\documentclass[aps,pra,onecolumn,preprintnumbers,10pt]{revtex4-2}

\usepackage{graphicx}  
\usepackage{subfigure}
\usepackage{multirow}
\Large

\usepackage[utf8]{inputenc}
\usepackage[english]{babel}

\usepackage{fancyhdr}
\usepackage{longtable}
\usepackage{parskip}
\usepackage[T1]{fontenc}
\usepackage{dcolumn}   

\usepackage{bm}        
\usepackage{amsfonts}  
\usepackage{amsmath}   
\usepackage{amssymb}   

\newcommand{\pwisein}{\left\{ \begin{array}{ll}}
\newcommand{\pwiseout}{\end{array}\right.}

\usepackage[unicode=true,pdfusetitle,
 bookmarks=true,bookmarksnumbered=false,bookmarksopen=false,
 breaklinks=false,pdfborder={0 0 0},pdfborderstyle={},backref=false,colorlinks=true]
 {hyperref}
\hypersetup{allcolors=blue}

\makeatletter
\renewcommand\NAT@citesuper[3]{\ifNAT@swa
\unskip\hspace{1\p@}\textsuperscript{[#1]}%
\if\relax#3\relax\else\ [#3]\fi\else [#1]\fi\endgroup}
\makeatother

\usepackage [english]{babel}
\usepackage [autostyle, english = american]{csquotes}
\MakeOuterQuote{"}

\begin{document}
\setcitestyle{super}
\title{Attochaos I: The classically chaotic postcursor of high harmonic generation}

\author{Jonathan Berkheim, David J. Tannor}

\affiliation {\it Department of Chemical and Biological Physics, Weizmann Institute of Science, 76100, Rehovot, Israel}


\begin{abstract}  
Attosecond physics provides unique insights into light-matter interaction on ultrafast time scales. Its core phenomenon, High Harmonic Generation (HHG), is often described by a classical recollision model, the simple-man or three-step model, where the atomic potential is disregarded. Many features are already well explained using this model; however, the simplicity of the model does not allow the possibility of classical chaotic motion. We show that beyond this model, classical chaotic motion does exist albeit on timescales that are generally longer than the first recollision time. Chaos is analyzed using tools from the theory of dynamical systems, such as Lyapunov exponents and stroboscopic maps. The calculations are done for a one-dimensional Coulomb potential subjected to a linearly polarized electric field.
\end{abstract}

\maketitle 

\section{Introduction}
\subsection{High harmonic generation and chaos}
Electron dynamics in atoms and molecules is typically on the timescale of attoseconds, $10^{-18}$s. For many years, there have been attempts to resolve electron dynamics and to capture the motion of charges directly within various media. Due to the ultrafast nature of the dynamics, only a “camera” with an attosecond frame rate can follow it adequately.\citep{Itatani2002AttosecondCamera} For this reason, High Harmonic Generation represents a unique capability for probing ultrafast electron dynamics.\citep{Corkum2007AttosecondScience,Krausz2009AttosecondPhysics}  High harmonics are generated when a strong electric field at a relatively low frequency, generally in the IR regime, impinges on an atom or molecule and results in the emission of a broad distribution of photons in the Extreme Ultra Violet (XUV) regime. This broad distribution in frequency corresponds, via Fourier synthesis, to an attosecond pulse train, which can then be used to capture the electronic motion.\citep{Hentschel2001AttosecondMetrology}

The properties of the emitted HHG photons cannot be explained within perturbative expansions. In fact, the high brightness of the emitted photons is indicative of the non-perturbative nature of the process, as first observed by L’Huillier.\citep{Ferray1988Multiple-harmonicGases} The higher-order responses to the induced field did not decay like a power law, as would have been expected on the basis of perturbation theory. A general time-dependent description for HHG is based on an atomic (Coulomb) potential subjected to an oscillatory electric field. The most commonly used field is linearly polarized, which drives the emission of bright harmonics at odd orders. Based on this description, Kulander et al. and Corkum introduced the approximate simple-man model, a three-step semiclassical model.

\begin{figure}[ht!]
     \centering
     \includegraphics[width=0.75\linewidth]{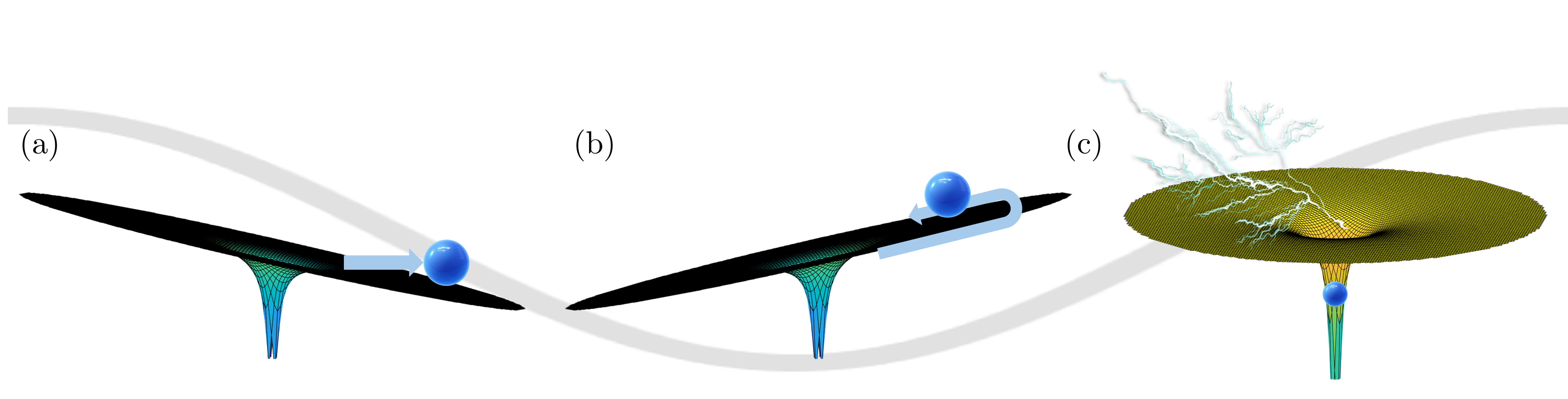}
     \caption{The three-step model simple-man model: (a) tunnel ionization under the lowered potential barrier (b) propagation (c) recollision and emission of a harmonic photon. The gray curve marks the oscillatory electric field. Adapted from \citep{Kern2015LimitationsNanophotonics}}
     \label{fig:three_step}
 \end{figure}

In the three-step model, the electron begins in the ground state of the Coulomb potential. \citep{Krause1993Super-IntensePhysics, Corkum1993PlasmaIonization} In the first step, a strong external AC electric field impinges on the atom and alternately lowers the Coulomb barrier so the electron can tunnel out at zero momentum, especially when the field is at its peak. The interplay between the field and the atomic potential displays an adiabatic character, quantified by the Keldysh parameter. In the second step of the model, the free portion of the electron is accelerated by the field and is described by Newtonian motion, i.e. the atomic force is neglected with respect to the force of the external electric field. In the third step, the field changes sign; the electron returns to the ion, where most of the electronic wavefunction has remained bound, and a recollision takes place. Most recollisions take place at nonzero momentum, and the excess kinetic energy of the electron is assumed to be emitted as photons having frequencies of integer multiples of the IR driving frequency; these are the HHG photons (see fig. \ref{fig:three_step}).

To quantify the second stage of the model, before the recollision event, one employs the classical equations of motion (EOM), where the ionization times are meant to reflect the instantaneous magnitude of the electric field. These equations are time-dependent and purely linear. Their solutions, i.e., the trajectories, do not exhibit any sensitivity to initial conditions; in other words, chaotic motion is not allowed in this model, at least not in one dimension, which is studied in this work. However, when the Coulomb potential is reintroduced the system acquires a new degree of complexity: the equations of motion will (1) contain a spatial potential independent of the field and (2) be nonlinear. We conjecture that chaos shall exist in this system, and we will show that this is the case indeed, using four independent tools from chaos theory: (1) maximal excursion in phase space (2) maximal Lyapunov exponent (3) stroboscopic maps (4) power spectrograms. With this understanding in hand, one can hope to explore possible signatures of chaotic dynamics in HHG spectra.

\subsection{A short review of previous works}
To the best of our knowledge, only a few previous works discussed the combination of chaos theory and high harmonic generation. Chism, Timberlake, and Reichl used Floquet theory to study classical chaos in particular systems of HHG, where the idealized potential describes either double resonance or a driven square well.\citep{Chism1998HighChaos} van de Sand and Rost presented semiclassical HHG spectra calculated with Gaussian wavepackets subjected to a Coulomb potential and a linearly polarized field; their main finding was that the formation of a plateau in the spectrum is mainly because of interference between irregular (i.e., chaotic) trajectories and regular orbits.\citep{vandeSand1999IrregularHarmonics} Fiordilino investigated HHG resulting from an artificial two-level chaotic system, where the states are coupled by a small nonlinearity and a dipole interaction.\citep{Fiordilino2016ChaosGeneration} Dubois et al. analyzed the phase-space representation of recollisions induced by a strong field in terms of invariant tori.\citep{Dubois2022DynamicalTori} Casati and Molinari explored the emergence of quantum chaos in a hydrogen atom in the presence of a strong electric field. In their work there is a single parameter - the magnitude of the electric field - and HHG is not mentioned at all.\citep{Casati1989QuantumHamiltonians} They found that classical chaos emerges if $\omega I_{p}^{3}\geq 1$, where $\omega$ is the frequency of the external electric field and $I_{p}$ is the ionization potential, namely the energy of the quantal ground state.

\subsection{The main innovations of this work
}This work aims to form a joint toolbox for HHG and chaos theory, which will help unravel deeper aspects of the corresponding classical dynamics that underlies HHG. In general, there is a major difference between characteristic timescales of HHG and chaos. On the one hand, the main features of harmonic spectra are formed quantum-mechanically within several optical cycles, and the repetitive recollisions at later times contribute to the narrowness of each harmonic peak. After a fairly long time, recollisions become weaker due to the spatial spreading of the electronic free wavepacket. From the classical point of view, one can capture the main spectral features by considering single-cycle dynamics. On the other hand, the dynamical character of a given classical trajectory, be it regular or chaotic, is determined after a relatively long time. One cannot predetermine whether a trajectory is chaotic or not by mere observation of its initial stages of motion.\citep{Heller2018TheSpectroscopy}

Thus, in order to properly investigate both features - recollision and chaos - we will carry out a “post-recollision analysis”, First, we will initiate an ensemble of classical trajectories with a variety of initial conditions. Then, we will select from among them the trajectories that recollide within the first cycle after ionization. We will let them continue evolving up to 100 optical cycles, orders of magnitude longer than the first recollision time. Finally, we will terminate the dynamics and use four criteria to determine whether chaos exists or not. The phenomenon of trajectories that are both recolliding and chaotic will be labeled as “attochaos”. Besides studying the dependence on the ionization times and the amplitude of the electric field, we also let the Coulomb potential vary in magnitude. This can be regarded as an additional parameter, which might give chaos a chance to emerge. The physical meaning of this varying Coulomb magnitude is to scan different chemical species. The work by Casati and Molinari, as well as several works on soft chaos in magnetic-field Hamiltonians,\citep{Friedrich1989TheChaos,Delande1989QuantumFields} did not let the Coulomb potential vary. Although this parameter is somewhat artificial from a physical perspective, we shall see that it adds a lot of richness from the point of view of dynamical systems theory and provides the bridge between the simple-man model and physical reality.

Below we rescale the equations of motion in order to explore the parameter space more effectively. Not only is this highly convenient, but it also makes the competition between two forces - Coulomb and external electric field - much more transparent. As mentioned above, the diagnostics of chaos that will be employed are various and independent of each other; this aims to ensure that the assigned dynamical character is indeed correct.

\section{Theory and Formulation}
\subsection{Rescaled equations of motion}
The full time-dependent Hamiltonian, beyond the simple-man model, is (in the length gauge)
\begin{equation}
\label{1.5D-Hamiltonian}
    H(x,p,t)=\frac{p^2}{2m}-qxE_{0}\sin\omega t-\frac{C}{\sqrt{x^2+a}},
\end{equation}
where $E_{0}$ is the field’s amplitude; $C,a$ are the parameters of the soft-core Coulomb potential, which avoids dealing with the singularity at the origin of the Coulomb potential; all other notations have their usual meaning. Implicit here is the assumption that the electron does not experience the spatial gradient of the electric field, thus $E(x,t)\approx E(t)$.\citep{Fleischer2007HighField} In the momentum gauge, the term $qx\cdot E(t)$ is replaced by the term $\frac{q}{c}p\cdot A$  where $A(t)=-c\int_{-\infty}^{t} E(t')dt'$  is the vector potential; however, Newtonian mechanics turns out to be invariant under such a transformation, then the equations of motion will look the same in both gauges.

A combination of Hamilton’s equations of motion yields Newton’s equation of motion:
\begin{equation}
\label{unscaled EOM}
m\frac{d^2}{dt^2} x=-qE_{0}\sin \omega t-\frac{Cx}{(x^2+a)^{3/2}}.
\end{equation}
We rescale this equation according to the following definitions
\begin{equation}
\label{scaling}
\tau\equiv\omega t,  \hspace{1cm} \tilde x\equiv x/\sqrt{a}, \hspace{1cm}  \tilde p\equiv \frac{d}{d\tau}\tilde x= p/m\omega \sqrt{a} ,
\end{equation}
and also employ atomic units, i.e., $m=|q|=1$. The reformulated form of eq. \ref{unscaled EOM} reads as follows
\begin{equation}
    \label{rescaled EOM}
    \frac{d^2 \tilde x}{d\tau^2}=-\tilde{E}_{0}\sin \tau-\frac{\tilde{C}\tilde x}{(\tilde x^2+1)^{3/2}},
\end{equation}
where we have defined
\begin{equation}
    \label{D1D2}
    \tilde{E}_{0}\equiv E_{0}/\sqrt{a}\omega^2,  \hspace{1cm} \tilde{C}\equiv C/a^{3/2}\omega^2,
\end{equation}
and we will also take $a=1$ for simplicity. In this process, we have reduced the number of parameters from $6$ ($q,m,a,E_{0}, C,\omega$) to $2$ ($\tilde E_{0},\tilde C$). Each selection of $\tilde E_{0}, \tilde C$ yields a “representative trajectory” which encapsulates the information on a family of trajectories. In other words, one can select an infinite number of combinations of $E_{0},C ,\omega$, that correspond to the same $\tilde E_{0}, \tilde C$; all trajectories will look the same, up to a scaling in phase space (in particular, as one can see from  \ref{scaling}, the positions will be identical, and the momenta will be scaled by $1/\omega$). 

\begin{figure}[ht!]
     \centering
     \includegraphics[width=0.3\linewidth]{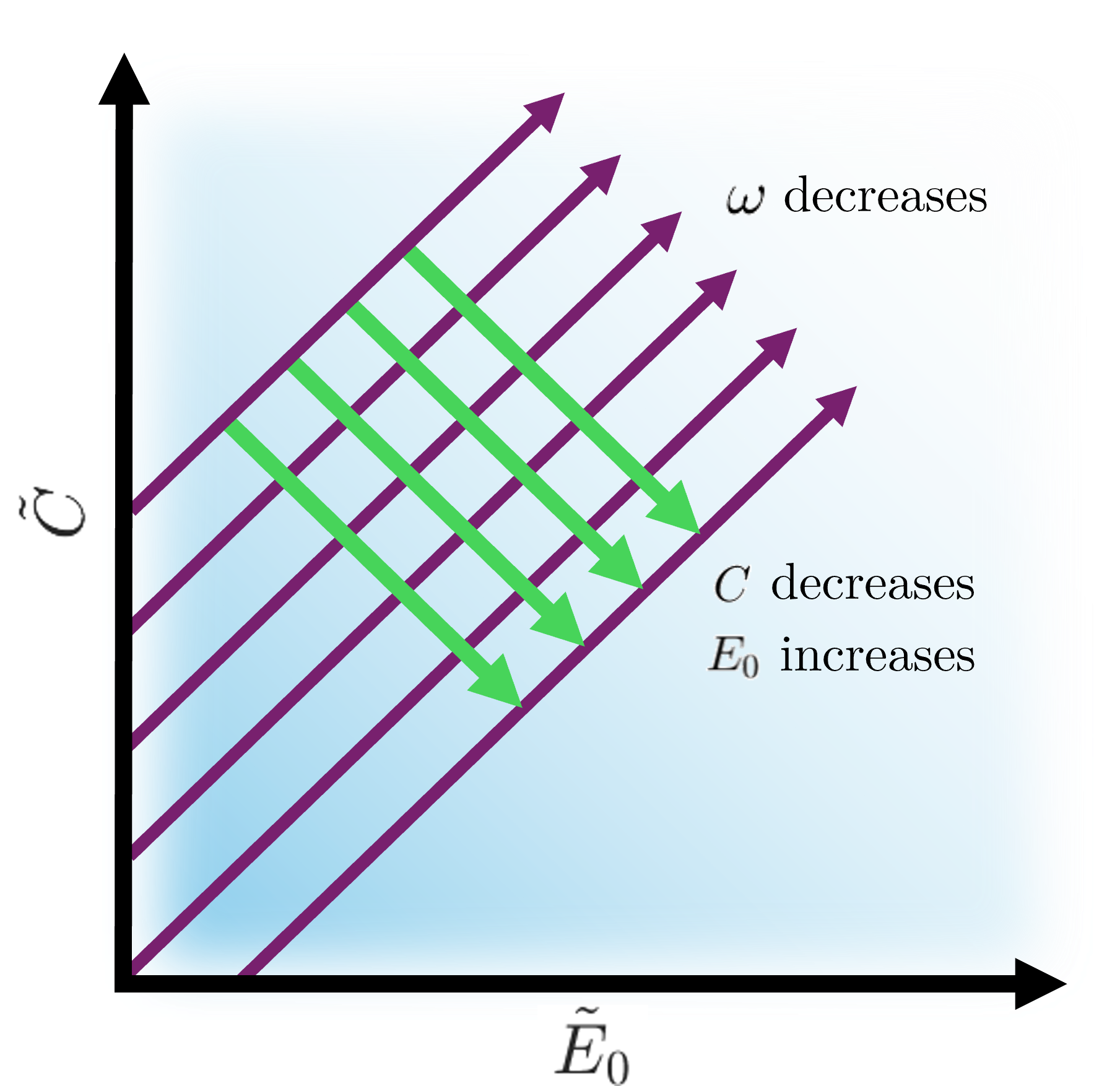}
     \caption{A visualization of a possible interpretation for the rescaled parameters.}
     \label{fig:geometrical_construction}
 \end{figure}

Experimentalists might find a helpful interpretation for the rescaled parameters in fig. \ref{fig:geometrical_construction}. We can examine two sets of diagonals: those where $\tilde C, \tilde E_{0}$ increase simultaneously (purple arrows), and those where $\tilde C$ decreases while $\tilde E_{0}$ increases (green arrows). In the first set, if we take fixed values of $E_{0},C$, then walking along the diagonals corresponds to decreasing $\omega$; in the second set, if we take fixed values of $\omega$, then walking along the diagonals corresponds to decreasing $C$ or increasing $E_{0}$. 

In the spirit of the simple-man model, we will initiate all trajectories from $\tilde x_{0}=\tilde p_{0}=0$, and scan over $\tilde{E}_{0}, \tilde{C}$ that both span from $0.01$ to $1$. The ionization times $ \tau_{i}$ shall also be provided as additional initial conditions and will span from $0$ to $0.5T$, with $T=2\pi$ being the rescaled optical cycle. The integration of eq. \ref{rescaled EOM} is done using the 4th-order Runge-Kutta algorithm,\citep{Runge1895UeberDifferentialgleichungen} and convergence is checked with respect to the time step $\Delta \tau$.

As shown in eq. \ref{rescaled EOM}, our system exhibits an interplay between two competing forces: the electric force and the Coulomb force, whose rescaled magnitudes are given by $\tilde{E}_{0}, \tilde{C}$. In general, these quantities contain the parameter $a$; if we take $a=1$, as said, then a particular combination of $\tilde{E}_{0}, \tilde{C}$ turns out to be almost identical to the celebrated Keldysh parameter,\citep{Popruzhenko2014KeldyshPerspectives}
\begin{equation}
\label{Keldysh}
\gamma=\sqrt\frac{I_{p}}{2U_{p}}\approx\frac{\sqrt{\tilde{C}}}{\tilde E_{0}},
\end{equation}
where $I_{p}\approx0.69C$ is the ionization potential of the atom (this relation holds only if $a=1$; in general $I_{p}=I_{p}(C,a)\propto C/\sqrt a)$, and $U_{p}=q^2E_{0}^{2}/4m\omega^2$ is the ponderomotive energy of an electron subjected to linearly polarized oscillating field (all in all, one gets a prefactor of $\sqrt{0.69\cdot2}=1.17\approx 1$). The function $I_{p}=I_{p}(C, a)$ is obtained by an imaginary-time relaxation of the time-dependent Schrödinger equation (TDSE),\citep{Kosloff1986AGrid} with a soft-Coulomb potential. The selected integration scheme is the 7-step split operator. \citep{BerkheimJ2022StudySpectroscopy}

In the case of a large Keldysh parameter, the Coulomb potential dominates over the electric field, and essentially, no HHG is assumed to expected; there are classical trajectories, but they have nothing to do with the three-step model. As we will see, they mainly express a bound motion, and their characteristic energies of recollision are extremely low; in this case, one can employ perturbative nonlinear optics and associate the spectral contribution of such trajectories with bound transitions, which appear in the spectrum as low harmonics.\citep{Fleischer2005DynamicalPulses}

In the case of an intermediate Keldysh parameter, where the Coulomb potential and electric field are comparable, many trajectories are expected to recollide. Some will exhibit regular behavior, and some will be chaotic, as we will show. These trajectories might be associated with Above-Threshold Ionization (ATI), a counterpart phenomenon of HHG, whose description exceeds perturbative optics.\citep{Fleischer2005DynamicalPulses,Cohen-Tannoudji2011AdvancesOverview}

The abovementioned time-dependent Hamiltonian is a generic classical description for an atomic system subjected to a strong field. Bear in mind that not all trajectories evolving under this Hamiltonian necessarily correspond to HHG; recollision depends on the interplay between the electric field and the Coulomb potential, which is expressed by the Keldysh parameter.

This work will concentrate on the contribution of classically chaotic trajectories to HHG, as many of them emerge at small Keldysh parameters, where the semiclassical model remains valid; nonetheless, we will also show some results obtained with large or intermediate Keldysh parameters, for the sake of completeness and in order to fully characterize the Hamiltonian.

\subsection{Regular vs. chaotic motion}
Nonlinear Hamiltonian systems are ubiquitous in nature. Such systems have been the subject of broad investigation over the last century since they appear in many real-life problems. 
Classical mechanics lives in a $2N$-dimensional phase space, where $N$ is the number of spatial degrees of freedom. In some instances, called quasi-periodic, regular, or integrable, there are $N$ conserved quantities so that the orbits live on an $N$-dimensional torus in the $2N$-dimensional phase space. For fully chaotic motion, the equations of motion are non-integrable; the torus structure hardly exists or does not exist at all.\citep{Gutzwiller1990ChaosMechanics} The only constant of motion in the time-independent case is the total energy, and in the time-dependent case, there are no conserved quantities unless one uses an extended phase space where time is an additional coordinate;\citep{Bensch1992EBKQuasi-energies,WustmannW2010StatisticalSystems} thus, the dynamics takes place in a $2N-1$-dimensional energy surface in phase space.

This profound geometrical difference between regular and chaotic motion underlies most tools that aim to distinguish between the dynamical behavior of trajectories; some of them will be employed in this work.

Time-dependent Hamiltonians, i.e., $H=H(t)$, are, in general, difficult to analyze since the energy is not conserved. Systems that include one degree of freedom and an explicit time dependence are sometimes referred to as “1.5D Hamiltonians”, as time is a parameter rather than a usual coordinate. The special case of time-periodic Hamiltonians, namely $H(t)=H(t+T)$ is formally treated by Floquet theory, a temporal analog of Bloch's theorem. One of the earliest uses of this theory was to find periodic orbits in Hill differential equation:
\begin{equation}
\label{hill}
\frac{d^{2}}{dt^2}{u}+f(t)u=0,
\end{equation}
with $f(t)$ a periodic function.

Any time-independent Hamiltonian system in one dimension is integrable; however, once the Hamiltonian is explicitly time-dependent (1.5D), classically chaotic motion might emerge. In fact, one of the simplest analytical models for chaos is a time-periodic Hamiltonian system called the Kicked Rotor (see the "future work" section).

The research on chaos in time-periodic systems began in the 1960s, when Contopoulos considered the integrals of motion for such case.\citep{Tzemos2021OrderSystems} Throughout the years, a fair number of works discussed the character of trajectories evolving under such Hamiltonians, and their relevance for, quantum chaos\citep{Casati1989QuantumHamiltonians} and solid state physics. In particular, the connection between the Kicked Rotor model and Anderson localization has been studied extensively.\citep{Fishman1982ChaosLocalization}

\subsection{Limiting cases}

We consider two limiting cases that can be integrated analytically. Rather than comparing $\tilde E_{0}$ to $\text{max}\left|\frac{\tilde{C}\tilde x}{(\tilde x^2+1)^{3/2}}\right|=\frac{2}{3\sqrt{3}}\tilde C\approx0.385\tilde C$, we will take an even strong conditions, where $\tilde E_{0}$ is compared to $\tilde{C}$.

In the first case, $\tilde{E}_{0}\ll \tilde{C}$, such that an approximate Hamiltonian will be
\begin{equation}
    \label{D1llD2_Ham}
    H\approx\frac{p^2}{2}-\frac{\tilde{C}}{\sqrt{\tilde x^2+1}},
\end{equation}
and, in accordance, Newton’s equation of motion can be approximated by
\begin{equation}
    \label{D1llD2_EOM}
    \frac{d^2 \tilde x}{d\tau^2}\approx-\frac{\tilde{C}\tilde x}{(\tilde x^2+1)^{3/2}},
\end{equation}
where the time-dependent term is neglected. In terms of the Keldysh parameter, this case corresponds to $\gamma \gg 1$, i.e., the multi-photon ionization regime. This force is short-ranged so, for simplicity, we consider here only $\tilde x\ll 1$, then Taylor-expand the force around the origin, up to first order in $\tilde x$:
\begin{equation}
    \label{taylor}
    \frac{d^2 \tilde x}{d\tau^2}\approx  -\tilde{C}\tilde x,
\end{equation}
and get an effective simple harmonic oscillator. The general solution will be
\begin{equation}
    \label{SHO}
\tilde x (\tau)=\tilde x_{0} \cos\left(\sqrt{\tilde{C}}\tau\right)+\tilde p_{0}\sin\left (\sqrt{\tilde{C}}\tau\right),
\end{equation}
along with its respective momentum $\tilde p (\tau)=\frac{d}{d\tau} \tilde x (\tau)$.

In the second case, $\tilde{E}_{0}\gg \tilde{C}$, such that an approximate Hamlitonian will be
\begin{equation}
    \label{timedep_Ham}
    H\approx\frac{p^2}{2}+\tilde x\tilde{E}_{0}\sin\tau,
\end{equation}
and, in accordance, Newton’s equation of motion can be approximated by
\begin{equation}
\label{timedep}
    \frac{d^2 \tilde x}{d\tau^2}\approx- \tilde{E}_{0} \sin \tau.
\end{equation}
In terms of the Keldysh parameter, this case corresponds to $\gamma \ll 1$, i.e., the tunnel ionization regime. The general solution will be
\begin{equation}
    \label{timdep sol}
    \tilde x(\tau)=\tilde x_{0}+\tilde p_{0}(\tau-\tau_{i})+\tilde{E}_{0}[\cos\tau-\cos\tau_{i}+(\tau-\tau_{i})\sin\tau_{i}],
\end{equation}
which, along with its respective momentum $\tilde p (\tau)=d\tilde x/d\tau$, forms an open, oscillatory curve in phase space; only in very particular cases, e.g., $\tau_{i}=0.25T$,  will there be a periodic solution, but these cases are of zero measure. Such trajectories do not exhibit chaos, as their Lyapunov exponents (see below) vanish; in other words, their stability is identical to that of a free particle. \citep{Heller2018TheSpectroscopy} 

It is obvious that in the first case, the Hamiltonian is symmetric, or invariant, under space-inversion and continuous time-translation $(\tilde x\mapsto-\tilde x,\hspace{0.25cm}\tau\to\tau+\epsilon,\hspace{0.25cm}\forall\epsilon)$ and the energy is conserved; in the second case, the Hamiltonian is symmetric under space-inversion and discrete time-translation $(\tilde x\mapsto-\tilde x,\hspace{0.25cm}\tau\mapsto\tau+T/2)$, which could be considered as a special case of the former, but in which the energy is not conserved. In the mixed regime, where neither $\tilde{E}_{0}\ll \tilde{C}$ nor $\tilde{E}_{0}\gg \tilde{C}$, these symmetries do not exist, and chaos might emerge. “Might emerge” rather than “must emerge” since the exact mixing of the competing forces might also give rise to regular motion, depending on the exact ionization time, $\tau_{i}$.

\section{Results and Analysis}

\subsection{Validity of the classical model}

A central tool in the classical analysis of HHG is the so-called an "attochirp curve",\citep{Krausz2009AttosecondPhysics} which can be interpreted as a classical spectrum. This curve, or more precisely the first half of it, displays the maximal recollision energies, $\mathcal{E}^*=\max[{\frac{1}{2} p^2(t_{r})}]$, where $t_{r}$ are the recollision times, against the ionization times $t_{i}$ (in this subsection we revert to the unscaled EOM). The information about recollisions is extracted from Newton’s equation in eq. \ref{unscaled EOM}, whose solutions $x(t)$ are numerically required to intersect (or to closely approach) the origin, i.e., $x(t_{r})\approx 0$. The second half of the attochirp curve, which is not discussed in this work, features the same maximal recollision energies against the recollision times.

efore delving into the chaotic features of the Hamiltonian, we would like to nail down the classical-quantum correspondence; in other words, we demonstrate that the full Hamiltonian successfully describes the quantal reality. 

From the set of maximal recollision energies, we can predict the set of emitted harmonic orders $n_{\text{HHG}}$ in the quantal spectrum   
\begin{equation}
\label{harmonic order}
n_{\text{HHG}}=\mathcal{E}^{*}/{\hbar \omega},
\end{equation}
due to the conservation of total energy in the light-matter interaction. Note that in the simple-man model, the numerator also includes the ionization potential $I_{p}$, which is added manually;\citep{Corkum1993PlasmaIonization} our Hamiltonian already contains the Coulomb potential, so there is no need to add a term of $I_{p}$.

In order to translate $\mathcal{E}^*$ to $n_{\text{HHG}}$, we shall select a particular frequency $\omega$ and a particular atomic potential. We take $\omega=0.05695\text{a.u.}=2.356\times10^{15}\text{Hz}$, which corresponds to $\lambda=800\text{nm}$, the central wavelength of Ti:Sa laser; the soft-Coulomb parameters are $C=0.02,a=1$, which lead to $I_{p}=0.09\text{a.u}=2.45\text{eV}$; the electric field amplitude is $E_{0}=0.08\text{a.u.}=4.11\text{V/m}$. All in all, this setup yields $\gamma=0.309$, such that a proper tunneling regime underlies the dynamics. 

\begin{figure}[ht!]
    \centering
    \includegraphics[width=0.7\linewidth]{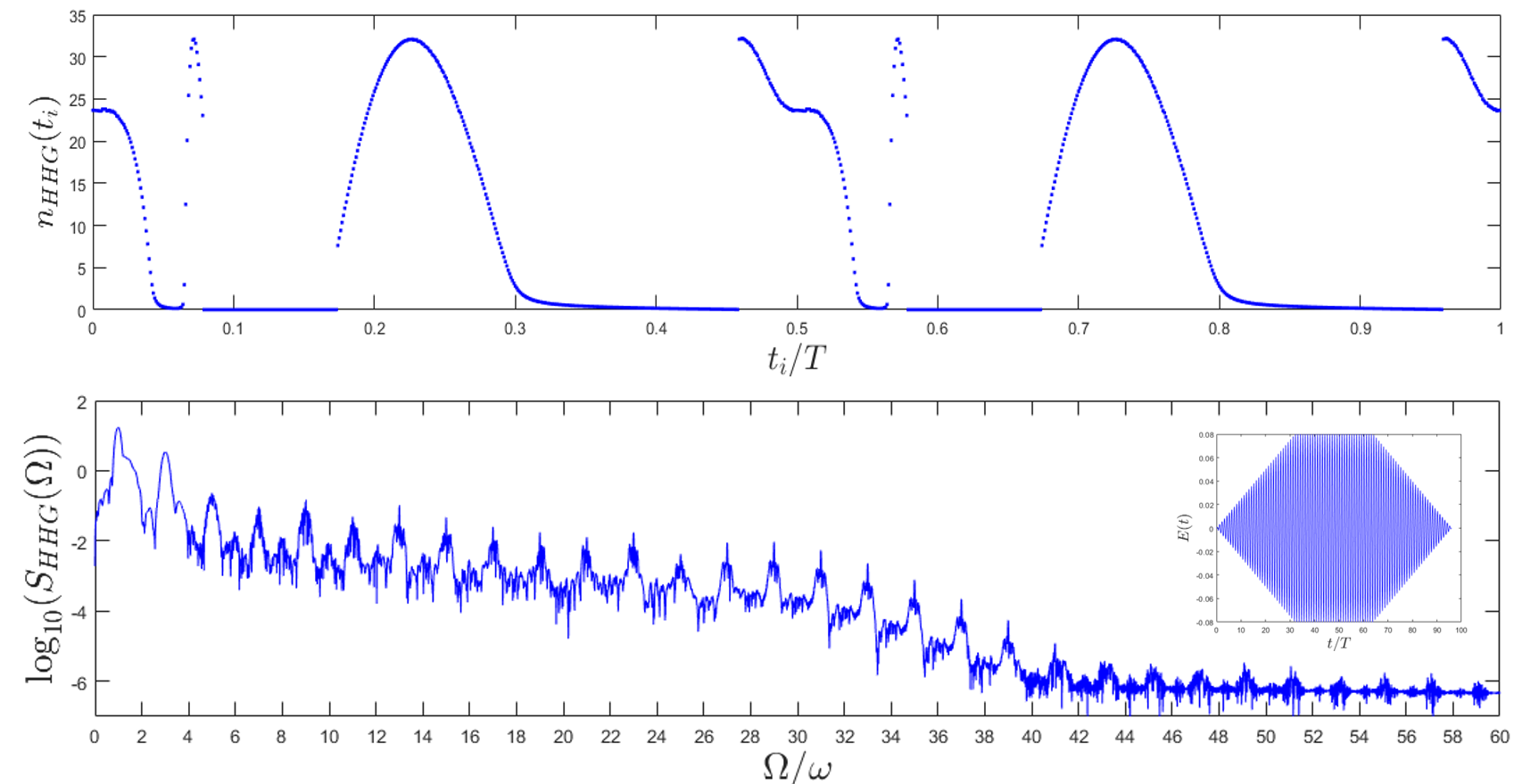}
    \caption{The classical attochirp curve, translated to harmonic orders, with $n_{\text{cutoff}}=32$ (upper plot); the quantal HHG spectrum, with odd harmonic orders and $n_{\text{cutoff}}=31$ (lower plot) and the induced electric field (inset). }
    \label{fig:Classical_quantum_correspondence}
\end{figure}

Scanning over various $t_{i}$ that span between $0$ and $T$, corresponding to a characteristic attochirp curve, the cutoff harmonic order $n_{\text{cutoff}}=32$ (see fig. \ref{fig:Classical_quantum_correspondence}; since even harmonics are not allowed for linearly-polarized driving field, we take $n_{\text{cutoff}}=31$ in practice. Note that due to the time-periodicity of the Hamiltonian, the region $0.5T<t_{i}<T$ is a perfect copy of the region $0<t_{i}<0.5T$. Henceforth, we will concentrate on $0<t_{i}<0.5T$.

We verify this classical prediction by solving a one-dimensional TDSE with the very same parameters, and subsequent calculation of the quantal HHG spectrum. The 1D grid is taken with $x_{\max}=300\text{a.u.}, \Delta x=0.25\text{a.u.}$, and the ground state is propagated up to $t_{f}=96T$; any unphysical contributions are avoided by absorbing boundary conditions around the grid ends. In the quantal simulation we take $E(t)=E_{0}f(t)\sin\omega t$ where $f(t)$ is a trapezoidal envelope reads as
\begin{equation}
    \label{trapez}
f(t)=\begin{cases}
t/32T, &  0\leq t/T<32 \\
1, &  32\leq t/T<64 \\
3-t/32T, &  64\leq t/T\leq96 \\
\end{cases},
\end{equation}
The HHG spectrum is calculated in the usual way of Fourier-transforming the expectation value of the acceleration\citep{Berkheim2023HighIonization}
\begin{equation}
\label{spectrum}
S_{\text{HHG}}(\Omega)=\left|\int_{0}^{t_{f}}\left\langle\Psi(t)\left|-\frac{1}{m}\frac{\partial V_{0}}{\partial x}\right|\Psi(t)\right\rangle e^{-i\Omega t}dt\right|,
\end{equation}
where $\Omega$ defines the frequency domain. We obtain a bright spectrum with harmonic cutoff  $n_{\text{cutoff}}=31$ (see fig. \ref{fig:Classical_quantum_correspondence}), where the spectrum starts decaying. With this classical-quantum correspondence in hand, a correspondence that is particularly convincing for the tunneling ionization regime, we can go forward with the rescaled model and scan different parameters.

\subsection{Recollision maps: classical spectra}

\begin{figure}[ht!]
    \centering
    \includegraphics[width=0.95\linewidth]{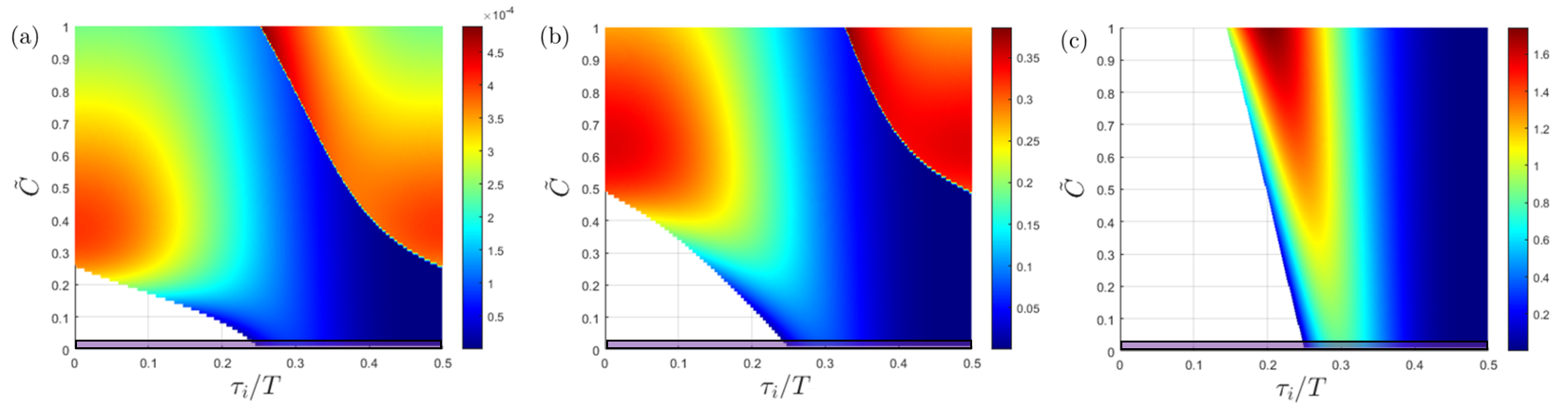}
    \caption{The maximal recollision energies $\mathcal{E}^{*}$ at various values of $\tilde E_{0}$: (a) $0.01$ (b) $0.31$ (c) $1$. (a) and (c) correspond to the limiting cases; (b) corresponds to the region of competition between $\tilde E_{0}$ and $\tilde C$. The white areas mean that no recollisions take place, and the lowest horizontal section in each subplot (marked by a framed purple bar) approximately recovers the 1D simple-man attochirp curve, where $\tilde C \to 0$.}
    \label{fig:attochirp}
\end{figure}

Now we return to the rescaled EOM, namely eq. \ref{rescaled EOM}. In fig. \ref{fig:attochirp}, we show recollision maps, or “2D attochirp curves”, rather than a single 1D attochirp curve. Here, $\mathcal{E}^{*}=\text{max}[\frac{1}{2}\tilde p^{2}(\tau_{r})]$ and the scanning is done on $\tilde C,\tilde E_{0}$and $\tau_{i}$. Such a 2D curve is a collection of 1D attochirp curves, each of them for a fixed value of $\tilde{E}_{0}$ and for a variable set of $\tilde{C}$. This way, one can study the effect of departure from the simple-man model on the maximal recollision energies; an attochirp curve for the simple-man model is approximately recovered in the lowest part of each recollision map, where $\tilde{C}\to 0$. Note that the most energetic recollisions are in the “simple-man forbidden regions”, i.e., the Coulomb attraction adds to the recollision energy.

For a fixed value of $\tilde E_{0}$, as $\tilde C$ increases, recollisions begin first at $\tau_{i}=0.25T$, and progress to $\tau_{i}=0$, namely, many new recollision channels in time are being accessed. For a fixed value of $\tilde C$, as $\tilde E_{0}$ increases, the system becomes closer to the simple-man regime, therefore the recollisions at $\tau_{i}<0.25T$ resemble the simple-man recollisions more and more (the original attochirp curve). Also, as $\tilde E_{0}$ decreases, the possibility of recollision grows but the energy of recollision decreases. Table \ref{tab:atto table} summarizes the main observations about these recollision maps.

Note that the time-periodicity of the Hamiltonian is implicit here: even though we took a reduced domain of $\tau_{i}$, each attochirp curve is periodic in $\tau_{i}$, namely $\mathcal{E}^{*}(\tilde C,\tau_{i})=\mathcal{E}^{*}(\tilde C,\tau_{i}+0.5T)$
\begin{table}[ht!]
    \centering
    \begin{tabular}{|c||c|c|c|c|} \hline  
           &$\tilde{E}_{0}$&Location of maximum in $(\tilde C,\tau_{i})$&  Dependence of $\mathcal{E}^*$ on $\tilde{C},\tau_{i}$& Deviation from the simple-man model\\ \hline  
           (a)&$0.01$& $0.39, 0.003T$&  strong, strong& many new recollision channels in time\\ \hline  
           (b)&$0.31$&$0.64, 0.004T$&  strong, moderate& fewer new recollision channels in time\\ \hline  
           (c)&$1$&$1, 0.2070T$&  strong, weak& almost no new recollision channels in time\\ \hline 
    \end{tabular}
    \caption{Recollision maps: main observations about fig. \ref{fig:attochirp}. Note the inverse correlation between field strength and (i) dependence on $\tau_{i}$ and (ii) extent of deviation from the simple-man model.}
    \label{tab:atto table}
\end{table}

\subsection{Criteria for chaos in various ionization regimes}
\subsubsection{Maximal excursion in phase space}
Maximal Lyapunov exponents (MLE) are widely considered good indicators for the existence of chaos,\citep{Heller2018TheSpectroscopy} and we will use them below as a quantitative criterion for chaos. However, before a detailed analysis based on MLE, it will be helpful to explore another qualitative introductory tool, the maximal excursion in phase space. First we define $D(\tau)$
\begin{equation}
    \label{Df}
    D(\tau)=\sqrt{[\tilde x(\tau)-\tilde x_{0}]^2+[\tilde p(\tau)-\tilde p_{0}]^2},
\end{equation}
An arbitrary but fairly large time will be used in this section for termination of the dynamics: $\tau_{f}=100T$ . In this interval, we will obtain the maximal $D$,  henceforth denoted $D_{\max}$. We conjecture that in regions associated with the limiting cases, there shall be relatively smooth changes in $D_{\max}$ under small variations in $\tilde{E}_{0}, \tilde{C},\tau_{i}$; in some of the regions associated with the mixed regime, there shall be relatively abrupt changes in $D_{\max}$ under small variations in $\tilde{E}_{0}, \tilde{C},\tau_{i}$. In particular, this can be seen as a Lyapunov-like indicator for sensitivity to the ionization time.

It is worth noting that the phase space distance is an important component of a quantitative chaos estimator called the pointwise dimension;\citep{Martens1984ADimension} this estimator is based on the abovementioned geometrical difference between regular and chaotic motion.

\subsubsection{Maximal Lyapunov exponent}
Lyapunov exponents quantify the rate of separation between infinitesimally close trajectories in phase space.\citep{Heller2018TheSpectroscopy} Each trajectory can be characterized by a maximal Lyapunov exponent
\begin{equation}
\label{Lyap}
\sigma=\lim_{\tau\to\infty}\lim_{\delta \vec \zeta_{0}\to 0}\frac{1}{\tau}\ln \left[\frac{|\delta \vec \zeta(\tau)| }{|\delta \vec\zeta_{0}|}\right],
\end{equation}
where $\delta\vec\zeta_{0}=(\delta \tilde p_{0},\delta \tilde x_{0}),\hspace{0.25cm}\delta\vec\zeta(\tau)=(\delta \tilde p(\tau),\delta \tilde x(\tau))$ are the initial and final separations (tangent vectors) in position and momentum for two close trajectories. A positive MLE is considered an indicator of chaos; otherwise, the trajectories are considered regular. Using this criterion, we can show analytically that the simple-man model yields regular trajectories:

In practice, the argument of the logarithm is replaced by a couple of eigenvalues $\lambda_{i}$ of the so-called stability matrix $\mathbb{M}$,\citep{Tannor2007IntroductionPerspective}
\begin{equation}
\mathbb{M}=\begin{pmatrix}
\frac {\partial \tilde p(\tau)}{\partial \tilde p_{0}} & \frac {\partial \tilde p(\tau)}{\partial \tilde x_{0}} \\
\frac{\partial \tilde x(\tau)}{\partial \tilde p_{0}}  &  \frac {\partial \tilde x(t)}{ \partial\tilde x_{0}}
\end{pmatrix}.
\end{equation}
In the case of the simple-man model,
\begin{equation}
\mathbb{M}=\begin{pmatrix}
1 & 0 \\
\tau-\tau_{i}  &  1
\end{pmatrix},
\end{equation}
where we inserted $\tilde x(\tau)$ from eq. \ref{timdep sol} and its respective momentum $\tilde p(\tau)$. The eigenvalues are $\lambda_{1,2}=1$ and in turn $\sigma=0$,  indicative of regular motion in the simple-man model.

In general, in numerical simulations we do not use the stability matrix formalism, and they might yield very small positive MLEs, such that chaos is suspected when the dynamics is actually regular. Thus, we will define an arbitrary threshold above which an MLE indicates chaos and under which it indicates regularity. At the current stage, we will take this threshold as $0.02$, a threshold that is justified by comparison with other indicators of chaos.

The initial separations are taken to be $1\times 10^{-6}$. The tangent vectors evolve in time by the linearized EOM,\citep{Littlejohn1992TheGeometry,Pikovsky2015LyapunovExponents} 
\begin{equation}
\delta \dot{\vec {\zeta}} = \mathbb{J}(t) \delta \vec{\zeta},
\end{equation}
where $\mathbb{J}$ is the Jacobian matrix,
\begin{equation}
\mathbb{J}=\begin{pmatrix}
-\frac {\partial^{2}H}{\partial \tilde x\partial \tilde p} & -\frac {\partial^{2}H}{\partial x^{2}} \\
\frac{\partial^{2}H}{\partial \tilde p^{2}}  &  \frac {\partial^{2}H}{\partial \tilde x\partial \tilde p}
\end{pmatrix}=\begin{pmatrix}
0 & \tilde C(2\tilde x^{2}-1)(\tilde{x}^{2}+1)^{-5/2} \\
1  &  0
\end{pmatrix}.
\end{equation}
We calculate only the MLE, obtained by the numerical Benettin algorithm:\citep{Benettin1980LyapunovTheory,Pikovsky2015LyapunovExponents} this simple routine is based on a normalization of the instantaneous tangent vector, such that divergences are avoided; the MLE is calculated as the average norm of the tangent vector,
\begin{equation}
\label{benettin}
\text{MLE}\approx \frac{1}{\tau_{f}}\sum_{n=1}^{N}||\delta\vec\zeta_{n}||,
\end{equation}
where $n$ enumerates the time steps such that $\tau_{i}=\tau_{1},\hspace{0.25cm} \tau_{f}=\tau_{N}$. In practice, early-dynamics transients are omitted, such that the summation starts at $\tau_{1}=T$.

\begin{figure}[ht!]
    \centering
    \includegraphics[width=0.95\linewidth]{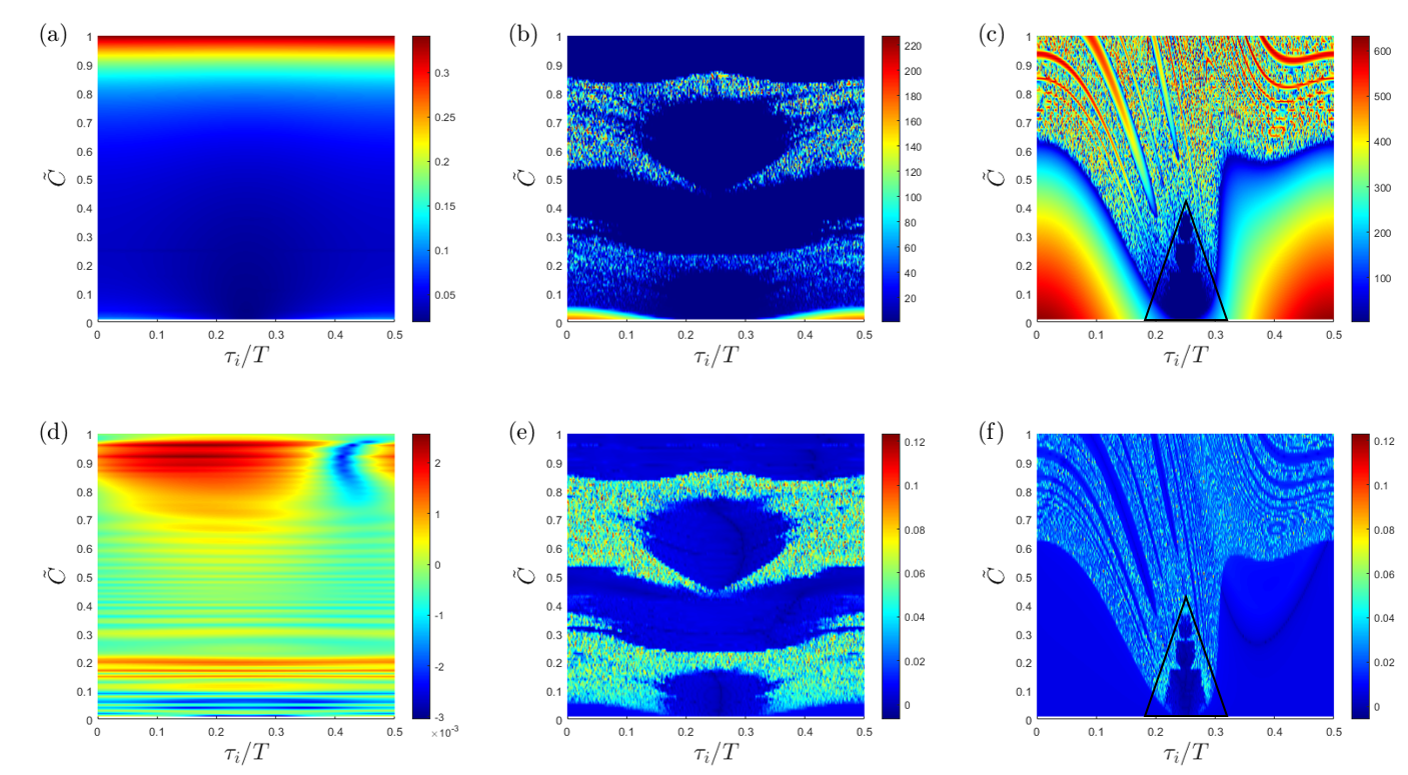}
    \caption{$D_{\max}$ (a-c) and MLE (d-e) at various ionization regimes: (a,d) $\tilde C=1, \tilde E_{0}=0.01$ (b,e) $\tilde C=0.83, \tilde E_{0}=0.31$ (c,f) $\tilde C=0.01, \tilde E_{0}=1$. The lowest horizontal section in each subplot approximately recovers the 1D simple-man, where $\tilde C \to 0$ and chaos does not emerge.}
    \label{fig:Dmax_and_MLE}
\end{figure}

In fig. \ref{fig:Dmax_and_MLE} we see a clear correspondence between the features of $D_{\max}$ and the counterpart MLE. Gradual changes in $D_{\max}$ (smooth phase) are associated with vanishing MLE and regular orbits; abrupt changes in $D_{\max}$ (granular phase) are associated with fairly positive MLE, thus indicating chaos.

There are many structural similarities between the maps of $D_{\max}$ and MLE: no chaos is predicted as $\tilde E_{0},\tilde C\to 0$;  the symmetry with respect to $\tau_{i}=0.25T$ at low and intermediate values of $\tilde E_{0}$; both maps exhibit a triangle shaped structure as $\tilde{E}_{0}\gg \tilde{C}$ (see (c,f) in fig. \ref{fig:Dmax_and_MLE}). It is not surprising, as both criteria encode the same dynamical information; in particular, the symmetry occurs due to an approximate invariance of the Hamiltonian under discrete time-translation. The triangle shaped hole in frames (c,f) in fig. \ref{fig:Dmax_and_MLE}) is associated with the first limiting case, where the electric force dominates over the Coulomb force. The triangle is centered around $\tau_{i}=0.25T$, an ionization time which gives rise to simple harmonic motion in the simple-man model, and where the electric field is at its peak. All the trajectories within this triangular region turn out to be regular. Even if we depart from the simple-man model, small $\tilde{C}$ keeps giving rise to regular orbits, reflecting the Kolmogorov-Arnol’d-Moser (KAM) theorem which guarantees the persistence of regular motion under small perturbations.\citep{Berry1978RegularMotion,Gutzwiller1990ChaosMechanics} We deduce that the triangular region defines the area within which the Coulomb potential can be considered a perturbation with respect to the electric field. Thus, its width decreases as we increase $\tilde{C}$.

\subsubsection{Stroboscopic maps}
A stroboscopic map is special conventional case of a Poincaré surface of section, appropriate for time-periodic Hamiltonians. In the conventional Poincaré surface of section for time-independent Hamiltonians, a reduced phase space (generically $p_{y}(y)$) is sampled at particular times when one of the dynamical variables, generically $x$, recurs to its initial value.\citep{Heller2018TheSpectroscopy}

Such surfaces provide the effective dimensionality of the trajectories in phase space: regular motion might look like a finite set of points (periodic orbits) or a well-ordered locus of points (quasi-periodic orbits); chaotic motion will seem like a sea of scattered points. These types of motion will foliate the allowed phase space, bounded by contours of constant energy. In time-periodic systems, there is no conservation of energy; however, one can exploit the time-translation symmetry of the Hamiltonian in order to examine the dynamical variables at integer numbers of periods, i.e., to plot $\tilde p(nT)$ vs. $\tilde x(nT)$, where $n\in\mathbf{N}_{0}$. In order to obtain a considerable distribution of points, we terminate the dynamics at longer times than before, $1000 T$; the same holds for the next section, where the resolution of the power spectrograms improves at longer times.

Due to space limitations, we will not show the stroboscopic maps for all possible trajectories; instead, we will focus on three particular sets of $\tilde{E}_{0}, \tilde{C}$ - one for each limiting case and the third for the mixed regime - and will show a collection of stroboscopic maps plotted on top of each other. Each map is associated with a particular $\tau_{i}$. This is not a conventional figure: normally one launches all trajectories at the same $\tau_{i}$ but different $(\tilde x_{0},\tilde p_{0})$; here we have $(\tilde x_{0},\tilde p_{0})=0$ for all trajectories, but different $\tau_{i}$. In conventional surfaces of section, invariant curves do not intersect each other, and there is an area preservation, but here none of these features is satisfied. 

The reason for deviating from the conventional stroboscopic format is that one can track how the parameter $\tau_{i}$ directly affects the dynamics for a given set of $\tilde E_{0},\tilde C$. We exploit this stacked representation in order to show a correlation between the attochirp curve and the stroboscopic maps: if we examine the vertical line where $\tilde{x}_{0}=0$, then we can read off the momentum of recollision at a particular time. This map supplements the attochirp curve, as it provides information about recollisions at integer times, which could, in general, be later than the short timescale of the attochirp. These recollisions are not necessarily the most energetic, yet it is still valuable additional information about later recollision times.

In all three cases, we take $\tau_{i}$ from $0$ to $0.5T$ at evenly-spaced steps (see the upper row of fig. \ref{fig:strobe_and_Fourier}). We easily distinguish regular and chaotic motion, according to the abovementioned thumb rules given for the Poincaré surface of section. In (a,c), all trajectories are regular; in (b), all trajectories are chaotic.

\subsubsection{Power spectrograms}
\begin{figure}[ht!]
    \centering
    \includegraphics[width=0.95\linewidth]{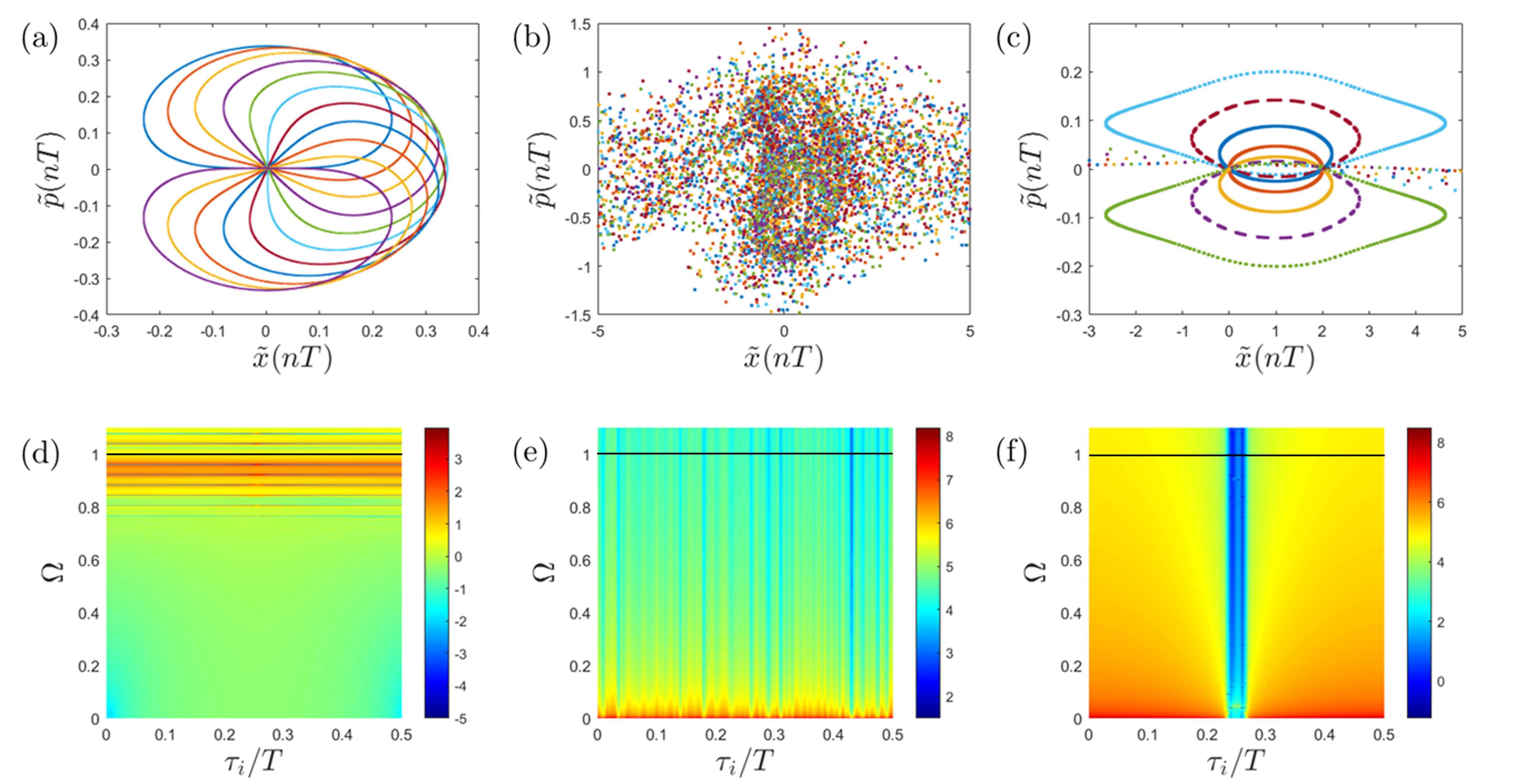}
    \caption{Stroboscopic maps (a-c) and their respective power spectrograms at particular sets of $\tilde{E}_{0}, \tilde{C}$: (a,d) $0.01,1$ (b,e) $0.31,0.83$ (c,f) $1,0.01$. Graph (b) captures a small area of the reduced phase space where the scattering of points is well observed.}
    \label{fig:strobe_and_Fourier}
\end{figure}

Numerous papers have claimed to recognize classically chaotic features in semiclassical spectra.\citep{Bixon1982QuantumSystem,Feit1984WaveSystem,Heller2018TheSpectroscopy,Tomsovic1991SemiclassicalAccuracy} It makes sense to conjecture that regular motion corresponds to a spectrum with well-defined, sharp peaks since the orbit is characterized by a finite set of frequencies. Chaotic motion, on the contrary, is expected to correspond to broad peaks,\citep{Stine1983MethodSystems} and rapid decay.\citep{Pomeau1986ChaoticPerturbation,Eidson1986QuantumField} Therefore, as another metric, we plot the classical power spectrograms of the same trajectories discussed in the previous subsection. Classical trajectories cannot yield an autocorrelation function unless given a finite width in phase space; however, we can still calculate their characteristic frequencies using the following Fourier transform 
\begin{equation}
\label{Dw squared}
 \mathcal{X}(\Omega)=\int \tilde x(\tau) e^{-i\Omega \tau}  d\tau.
\end{equation}
All spectrograms shown in the lower row of fig. \ref{fig:strobe_and_Fourier} display a sharp peak at $\Omega=1$ (black lines were marked manually for convenience), which corresponds to the field’s effective frequency, i.e., $E(\tau)=\sin1\tau$. All trajectories in the first limiting case ($\tilde{E}_{0}\ll \tilde{C}$) and some of the trajectories in the second limiting case ($\tilde{E}_{0}\gg \tilde{C}$) are represented by clean spectra with sharp peaks. In contrast, the mixed regime, which features fully chaotic behavior with these parameters, is characterized by rapidly decaying spectra. One might suspect that many trajectories in the mixed regime and in the second limiting case have the same spectrum but in fact these two sets differ in their behavior. In the second limiting case, there are gradual changes in the intensity profile under a variation of $\tau_{i}$; these changes are associated with the smooth region of the $D_{\max}$ plot. In the mixed regime, there are mostly abrupt changes (vertical lines) in the intensity profile under a variation of $\tau_{i}$, similar to what we had in the corresponding $D_{\max}$ plot.

\subsection{Chaotic fractions}
Here we provide an estimate of the relative fraction of chaotic trajectories for a particular selection of parameters. Note that many combination of parameters do not correspond to HHG. Each selection of $\tilde{E}_{0}, \tilde{C}$ is accompanied by a scan over a set of $\tau_{i}$; in many cases, not all of the $\tau_{i}$ give rise to a single type of motion, regular or chaotic. For this reason, one can characterize particular selections simply by counting the number of chaotic trajectories among the entire set of trajectories obtained with this selection. The criterion for chaotic motion, in this case, will be a threshold value for the MLE, taken to be $0.02$, as before. This way, we define the parametric chaotic fraction, denoted $\mu$,
\begin{equation}
\label{chaotic fraction}
\mu(\tilde{E}_{0},\tilde{C})=\sum_{k=1}^{N}\left\{ \begin{array}{cc}
1/N, & \text{MLE}_{k}\geq0.02\\
0, & \text{MLE}_{k}<0.02
\end{array}\right\},
\end{equation}
where $k$ enumerates the set of ionization times taken per each selection of $\tilde{E}_{0}, \tilde{C}$. A similar procedure was tried in the past for the system of a hydrogen atom subjected to a magnetic field.\citep{Friedrich1989TheChaos} The summation in eq. \ref{chaotic fraction} removes the dependence on $\tau_{i}$ and quantifies, in a single number per each selection, the percentage of chaotic trajectories.

In the same fashion, we define the temporal chaotic fraction $\nu$, in which the dependence of the dynamical behavior is integrated over different selections of parameters, such that only $\tau_{i}$ is left
\begin{equation}
\label{temp chaotic fraction}
\nu(\tau_{i})=\sum_{i,j=1}^{N}\left\{ \begin{array}{cc}
1/N, & \text{MLE}_{i,j}\geq0.02\\
0, & \text{MLE}_{i,j}<0.02
\end{array}\right\},
\end{equation}
where $i,j$ enumerate the sets of $\tilde E_{0},\tilde C$. The key idea was to see whether there are any preferred ionization times that give rise to chaos.

\begin{figure}[ht!]
    \centering
    
    \includegraphics[width=0.65\linewidth]{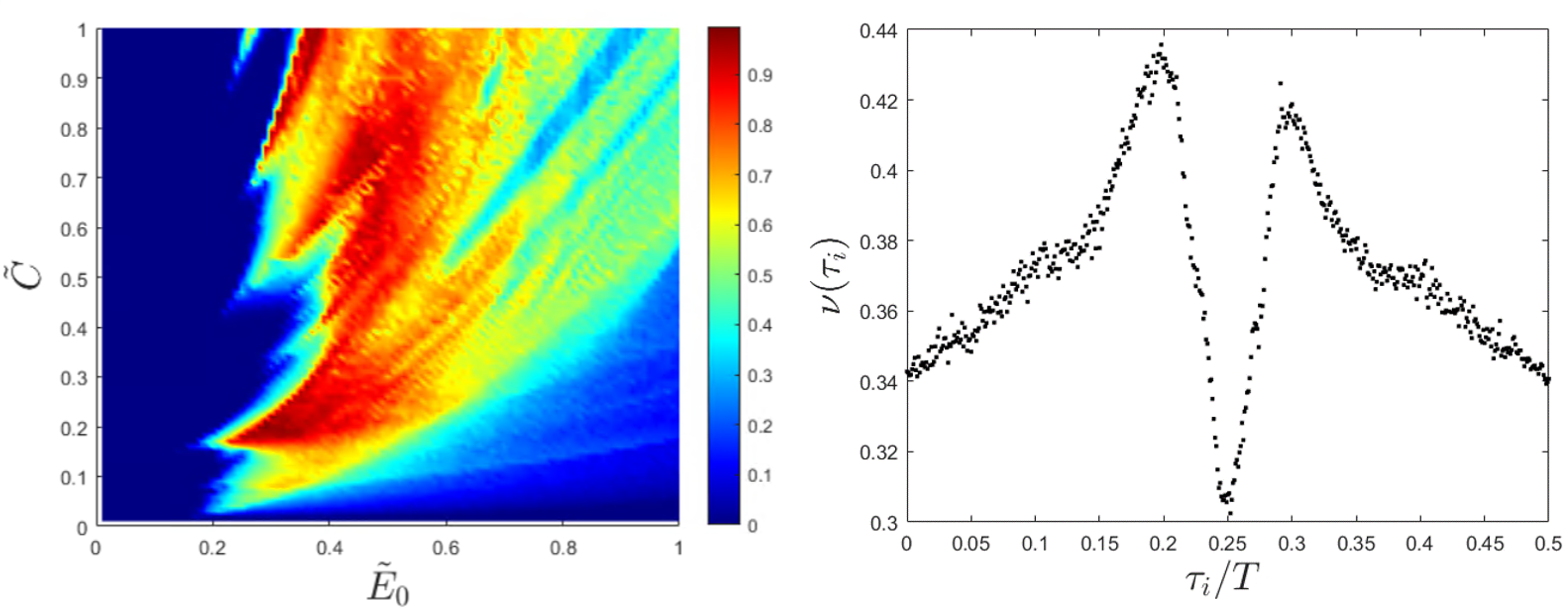}
    \caption{The transition to chaos is depicted by the continuous changes in the parametric chaotic fraction $\mu$ (left) and temporal chaotic fraction $\nu$ (right). $\mu$ exhibits low values in the limiting cases, and $\nu$ is at a minimum around the simple harmonic motion in the simple-man case.}  
    \label{fig:phase_transition}
\end{figure}

The surface plotted in fig. \ref{fig:phase_transition} shows $\mu(\tilde{E}_{0}, \tilde{C})$. It is evident that in the limiting cases, $\mu\approx0$, and in the mixed regime, where $\tilde{E}_{0}, \tilde{C}$ are of the same order of magnitude, there is a chance for chaos, implying a coexistence of regular and chaotic trajectories. In one case 
($\tilde{E}_{0}=0.31,\tilde{C}=0.83$ and $\gamma=2.93$) we obtain exactly $\mu=1$, which implies that all involved trajectories are chaotic (taking the abovementioned threshold for the MLE) . This is the reason why we focused on $\tilde{E}_{0}=0.31$ in the previous sections. Note that for $\tilde{E}_{0}=\tilde{C}=0.5$ (and, $\gamma=1.41$),  $\mu<1$, not $\mu=1$ as might have been expected. 

Regarding the temporal chaotic fraction, we clearly see that $0.3\leq\nu(\tau_{i})\leq 0.44$.  This means that each $\tau_{i}$ leads in general to a coexistance of regular and chaotic trajectories. Around $\tau_{i}=0.25T$, there emerges a global minimum; this corresponds to the fractal structure in fig. \ref{fig:Dmax_and_MLE}, where there is a persistent fraction of regular trajectories at various selections of parameters. 
As one departs even slightly from $0.25T$, the fractal structure is replaced by a large concentration of chaotic trajectories. This is most prominent at intermediate and large values of $\tilde E_{0}$, and accordingly $\nu$ will be maximum there. Larger deviations from $0.25T$ will yield a larger fraction of regular trajectories, such that $\nu$ will be smaller.

\subsection{Attochaos and strong chaotic recollisions}
At this stage, we would like to combine the information about recollisions and chaotic motion. Not every recolliding trajectory is chaotic, and not every chaotic trajectory is recolliding;  also, the timescales when energetic recollisions and chaos are measured are very different. To find those trajectories that possess both features, we will now filter out trajectories that are both strongly recolliding and chaotic. The threshold energy for strong recollision will be $\mathcal{E}^*\geq0.5$ (by eq. \ref{harmonic order}, this corresponds to $n_{\text{HHG}}\geq9$ if $\omega=0.05695\text{a.u.}$); chaotic motion will be determined by an MLE of $0.02$ or larger, as before.

\begin{figure}[ht!]
    \centering
    \includegraphics[width=0.35\linewidth]{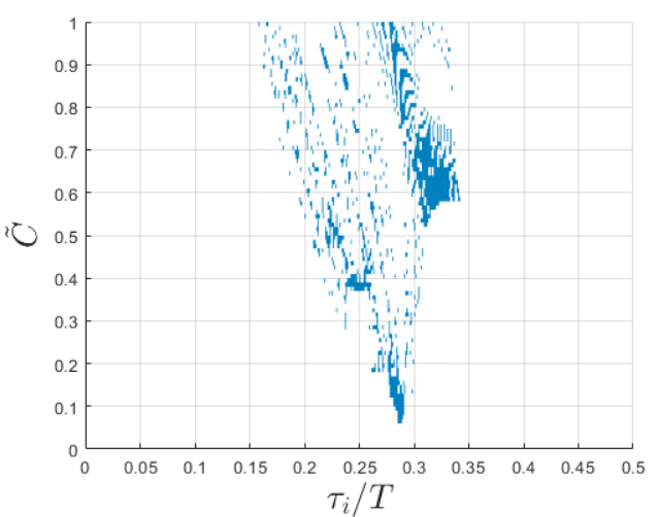}
    \caption{A map of strong chaotic recollisions obtained for $\tilde{E}_{0}=1$. Note that only trajectories obtained with $\gamma<1$ deserve the labeling "attochaos".}
    \label{fig:attochaotic_trajectories}
\end{figure}

A collection of attochaos trajectories, namely, those that satisfy both criteria and have $\gamma<1$, are shown in fig. \ref{fig:attochaotic_trajectories}. The units of these maps are arbitrary, as they depict an overlap between $\mathcal{E}^*$ and MLE, two quantities whose product is physically meaningless; for this reason, we plot a binary map, where “attochaotic” trajectories are signaled by “$1$”; otherwise, “$0$”. Not surprisingly, attochaos does not exist in the limiting cases, either because there is no chaotic motion there or because the recollisions are too weak. 

We cannot say that $\tilde{E}_{0}\leq \tilde{C}$ or $\tilde{E}_{0}\geq \tilde{C}$  are necessarily demanded for strong chaotic recollisions; both relations might work. However, we can make the blanket statement that chaos is unlikely to emerge in the limiting cases, as seen in fig. \ref{fig:attochaotic_trajectories}. The associated Keldysh parameters as calculated by eq. \ref{Keldysh}, are $0.24\leq\gamma\leq2.5$, i.e., all three ionization regimes may exhibit chaos, but in order to relate these trajectories with HHG, we need $\gamma<1$.
\section{Conclusion}
In this work, we have established a basic joint language for combining the field of attosecond physics with chaos theory. As described, this is not the first work that discusses possible relations between these two fields; however, by writing down a rescaled version of the equations of motion based on merely two controllable parameters, we were able to scan over an extensive space of combinations that were not discussed in previous works. Employing and combining four independent metrics, we proved that chaotic motion is definitely involved in the classical manifestation of high harmonic generation and that strong chaotic recollisions might take place in a variety of ionization regimes, although not in the limiting cases. We found that chaos emerges in the mixed regime of ionization, where the Hamiltonian does not possess any symmetry; certain trajectories in the chaotic regions are also strongly recolliding and can therefore be associated with the emission of high harmonic radiation. 

We hope that this work will stimulate further works at the intersection between chaos theory and attosecond physics.

\section{Future Work}

\subsection{From simple-man to kicked rotor}

As discussed in the text, trajectories evolving as free particles with a driving field of just a single frequency, e.g., $E(t)=E_{0}\cos\omega t$, do not exhibit chaos. On the other hand, there is a well-studied model for chaos in 1.5D dynamics called the Kicked Rotor described by the Hamiltonian
\begin{equation}
\label{KR}
    H=\frac{L^{2}}{2I}+K\cos\theta \sum_{n=-\infty}^{\infty}\delta(t-nT),
\end{equation}
where $K$ is the kick's amplitude, $I$ is the moment of inertia, $L$ is the angular position of the rotor, and $T$ is the kick periodicity. The kicked rotor is appealing because the dynamics can be transformed to a recursive map relating the position and momentum at each time step with that of the previous time step. Comparing the kicked rotor with the simple-man Hamiltonian
\begin{equation}
\label{HHG-KR}
    H=\frac{p^{2}}{2m}+E_{0}x \cos\omega t,
\end{equation}
we note the following correspondence: $m\to I,\hspace{0.25cm}E_{0}\to K,\hspace{0.25cm} x\to \cos \theta,\hspace{0.25cm} L\to p$  and $\cos \omega t \to \sum_{n=-\infty}^{\infty}\delta(t-nT)$. We raise the question: why is there chaos in the kicked rotor and not in the simple-man model? Is it due to the vs. $x$ factor or to the train of pulses $\sum_{n=-\infty}^{\infty}\delta(t-nT)$ vs. the single frequency $\cos \omega t$? It turns out that the key difference is the latter.\citep{R.Blumel1997ChaosPhysics} We can go deeper with the comparison by noting that the pulse train in the kicked rotor can be represented as an infinite sum over all harmonic orders: 
\begin{equation}
\label{ATP}
\sum_{n=-\infty}^{\infty}\delta(t-nT)=\sum_{n=-\infty}^{\infty}e^{in\omega t}=2\sum_{n=1}^{\infty}\cos n\omega t+1,
\end{equation}
i.e., as an infinite set of harmonic frequencies including a DC component. This raises several additional questions: 1) How many frequencies are necessary to begin to observe chaos in the kicked rotor? 2) How important is the DC component in producing chaos in the kicked rotor? Not much is understood about this transition \citep{Masovic2021KickedTrain,R.Blumel1997ChaosPhysics}and we intend to study this more fully in future work. If only the DC component is present, the Hamiltonian will be
\begin{equation}
\label{KR-only-DC}
H=\frac{L^{2}}{2I}+K\cos\theta,
\end{equation}
which is the Hamiltonian of a pendulum. This is an integrable system, analogous to the time-independent Hamiltonian in eq. \ref{D1llD2_Ham}. If the DC component is replaced by a single AC component, the Hamiltonian will be 
\begin{equation}
\label{KR-only-AC}
H=\frac{L^{2}}{2I}+2K\cos\theta\cos\omega t,
\end{equation}
which is equivalent to the Hamiltonian of a free particle driven by an electric field. This should also be an integrable system, analogous to the time-dependent Hamiltonian in eq. \ref{timedep_Ham}.  However, once we combine the DC component with even a single AC component, namely
\begin{equation}
\label{KR-ACDC}
H=\frac{L^{2}}{2I}+K\cos\theta+2K\cos\theta\cos\omega t,
\end{equation}
the Hamiltonian deviates from the simple-man model, and is analogous to our full Hamiltonian (eq. \ref{1.5D-Hamiltonian}). Thus, we conjecture that without the DC component, this system will not be chaotic; with the DC component, even with one driving field it could be chaotic.

\subsection{Initial momentum and higher-dimensions driving schemes}

As noted above, a linearly polarized electric field produces only odd harmonics due to symmetry restrictions. The emitted harmonics will be linearly polarized as well due to the same restrictions. There have been attempts to generate circularly polarized harmonic radiation using a circularly polarized driving field in a two-dimensional setup. However, this is impossible due to the violation of symmetry restrictions.

To overcome these symmetry restrictions, the notion of two-color counter-rotating driving fields was introduced, theoretically\citep{Milosevic2000GenerationMixing} and experimentally.\citep{Fleischer2014SpinGeneration}  This two-color scheme allows the emission of high harmonic photons; if almost equal numbers of incident photons have right- or left-circular polarization, then one finds new conservation rules regarding the polarization and energy. The most common two-color scheme combines circularly polarized fields at frequencies $\omega$ and $2\omega$, i.e., fundamental and Second Harmonic Generation (SHG) photons with opposite helicities
\begin{equation}
\vec E(t)=\frac {E_{0}}{\sqrt 2}
\begin{pmatrix}
\cos \omega t + \cos 2\omega t  \\ \sin \omega t - \sin 2\omega t
\end{pmatrix},
\end{equation}
a field whose wavefront possesses a 3-fold symmetry. It can be shown that each $3n$ order is absent in the HHG spectrum for all ${n\in \mathbf{N}}$, and all the allowed orders will be circularly polarized with alternating helicity. The brightness of such a spectrum is comparable to that obtained with a linear field; however, the success of the two-color counter-rotating scheme is somewhat mysterious. The use of two colors overcomes the symmetry restriction on HHG for circular polarization, however under the simple-man model there is no recollision unless a nonzero initial momentum is provided. Although the simple-man model assumes that ionization begins with zero momentum, we can associate a nonzero momentum with quantum considerations: (1) As the electron tunnels through the lowered potential barrier, it is highly localized in position, and, due to Heisenberg’s Uncertainty Principle, the wavefunction it is highly delocalized in momentum space (2) we can apply an additional XUV field to form an initial scattering state with nonzero momentum. This variation on the three-step model has been previously considered.\citep{Fleischer2008GenerationModel} 

Nonzero initial momentum modifies the subsequent electron momentum and position and the entire attochirp curve, as the electron can miss the ion or recollide at higher energy, depending on the interplay between $t_{i}$ and the initial conditions. We conjecture that it will also affect all the metrics used here to determine chaotic motion. It will be interesting to check different initial momenta and their influence on the chaotic fraction and the attochaos maps. Although the effect of initial momentum on attochaos in systems with two helicities (and therefore two dimensions) is intriguing (additional dimensions might imply additional routes to chaos due to the coupling between $x,y$) it will be worthwhile to investigate these effects in one dimension as well. Recent work also discussed the effect of initial momentum on HHG driven by a co-rotating two-color field.\citep{Berkheim2023HighIonization}

\break
\section{{Acknowledgements}}
Financial support for this work was provided by the Israel Science Foundation (1094/16 and 1404/21), the German-Israeli Foundation for Scientific Research and Development (GIF), and the historic generosity of the Harold Perlman family. JB and DJT thank Vered Rom Kedar, Nirit Dudovich, Yair Zarmi, Omer Kneller and Barry Bruner for helpful discussions. 

\bibliographystyle{apsrev4-2}

\end{document}